\documentclass[aps,prapplied,reprint,superscriptaddress]{revtex4-2}
\usepackage{graphicx}
\usepackage{dcolumn}
\usepackage{bm}
\usepackage{physics}
\usepackage{comment}
\usepackage{algpseudocode}
\graphicspath{{./figures/}}
\usepackage{color}
\usepackage{xcolor}
\begin{document}
	\title{Influence of Point Defects on Laser-Induced Excitation in Silicon}
	\author{Tomohito Otobe}%
	\email{otobe.tomohito@qst.go.jp}
	\affiliation{%
		Kansai Institute for Photon Science, National Institutes for Quantum Science and Technology (QST), Kyoto 619-0215, Japan
	}%
	\affiliation{%
		Photon Science Center, Graduate School of Engineering, The University of Tokyo, 7-3-1 Hongo, Bunkyo-ku, Tokyo 113-8656, Japan
	}
	
	\author{Eiyu Gushiken}
	\affiliation{%
		Department of Nuclear Engineering and Management, Graduate School of Engineering, The University of Tokyo, 7-3-1 Hongo, Bunkyo-ku, Tokyo 113-8656, Japan
	}%
	\begin{abstract}
		We studied the influence of defect states on the laser excitation process in silicon using time-dependent density functional theory. We assumed two types of point defects: interstitial oxygen and silicon vacancies. We found that the photoabsorption efficiency increased with defect density in both cases owing to the color center. These defects distorted the crystal structure, thereby relaxing the selection rules and changing the indirect gap to direct. 
		At low laser intensities, the defect states dominated the absorption process. However, as the laser intensity increased, the excitation efficiency approached that of crystalline silicon. 
		In addition, we observed that the excitation efficiency did not scale linearly with the pulse length. Notably, in the case of Si vacancies, saturable absorption significantly reduced photoabsorption.
		Our results suggest that the existence of a defect induces the even-order high-harmonics generation. The enhancement of even-order high-harmonics generation by the silicon vacancy is larger than the interstitial oxygen. 
	\end{abstract}
	\pacs{Valid PACS appear here}
	
	\maketitle
	
	\section{Introduction}
	In recent years, the use of intense laser pulses with durations ranging from femtoseconds to picoseconds has significantly advanced our understanding of laser-matter interactions. Ultrashort laser pulses are particularly important for the precise processing of gapped materials, as they allow nanostructuring without damaging the surrounding areas \citep{Pronko1998,Tien1999, Gattass:2008aa,Phillips2015,Stoian2020,Bonse2020}. This precise control is essential for nanotechnology and materials science applications \cite{Sugioka:2014aa}. Laser processing encompasses a broad range of physics, including optics, quantum mechanics, material science, and thermodynamics, necessitating robust numerical modeling to comprehensively optimize and understand these processes. Over the years, the numerical modeling of laser excitation has progressed significantly\citep{Agassi1984,Driel1987,Chen2005,Popov2011,Lipp2014,Tsibidis2012,Venkat_2022}, with silicon being a material of particular interest because of its fundamental scientific and industrial applications \cite{Sokolowski1995,Sokolowski1998}.
	
	While theoretical studies have focused on single-shot processes with perfect crystals, the complexities of multi-pulse processing with various target qualities remain less explored. Multi-pulse processing introduces cumulative effects, such as temperature changes, structural modifications, and melting \cite{Bonse2002,Izawa2006,Ito2018}. Recently, burst-mode processing has attracted interest because of its efficiency in exciting materials before thermal diffusion occurs \cite{Kerse:2016aa}. During multi-shot irradiation, increased target temperatures can induce structural changes, including self-trapped excitons \cite{Mero2005}, surface morphology alterations, and fluid dynamics in molten materials \cite{Bonse2020}. Moreover, the quality of the target material, including doping, domain structures, defects, and composite structures, is crucial in laser processing. New computational methods that incorporate diverse material properties are required to accurately model laser processing in such complex materials. Owing to the complexity of non-perturbative effects, first-principles simulations are ideal for estimating laser excitation efficiency. Among these, time-dependent density functional theory (TDDFT) \cite{TDDFT1984} is a highly accurate approach with a feasible computational cost, which has been successfully applied to nonperturbative laser excitation processes, contributing to our understanding of laser-induced damage \cite{Otobe2008,Yabana2012,Sato2015,Otobe2019,otobe2019JAP,Otobe2020,Otobe2020APL}. To simulate laser-matter interactions, we have developed the scalable ab initio light–matter simulator for Optics and Nanoscience (SALMON) code, which is based on TDDFT \cite{SALMON}.

	In this study, we investigate the excitation of silicon in the presence of point defects. Point defects are crucial for both accumulation effects without noticeable structural changes and in the overall vulnerability of the material. The electronic structures of silicon with defects can be accurately calculated using first-principles approaches such as density functional theory (DFT) \cite{Pesola1999}. To fully understand the impact of defects on laser processing, their effects in nonlinear and non-perturbative regimes must be explored. 
	
	Our study used the SALMON code to focus on electron dynamics in silicon with typical point defects, specifically interstitial oxygen (O$_i$) and silicon vacancies (V$_{Si}$). This advanced simulation tool allowed us to model the complex interactions between intense laser fields and defect-laden silicon, providing insights into the underlying mechanisms of laser excitation and potential pathways for optimizing laser processing techniques.
	
	The remainder of this paper is organized as follows: Sec.~\ref{sec:method} describes the
	theoretical methods and numerical simulations  
	In Sec.~\ref{sec:results},
	the calculation results are presented and analyzed in detail.
	Finally, the conclusions are presented in Sec. ~\ref{sec:summary}.
	
	\section{Computational method \label{sec:method}}
	In the first step, we introduced point defects at arbitrary positions in a $4 \times 4 \times 4$ supercell containing 512 Si atoms. In the second step, we used DFT with Quantum ESPRESSO (QE) \cite{QE1,QE2} to calculate the relaxed atomic structure, including the defects. Finally, the electron dynamics were calculated using SALMON.
	
	For the atomic structure calculations, we used the pseudopotentials Si.pbesol-nl-rrjkus\_psl.1.0.0.UPF and O.pbesol-n-rrkjus\_PSL.1.0.0.UPF from the QE pseudopotential database (http://www.quantum-espresso.org/pseudopotentials), along with the PBEsol exchange-correlation potential \cite{PBEsol2008}. The specific parameters used in the calculations are listed in Table \ref{t1}. 
	
	We employed the SALMON code for real-time electron dynamics calculations. The details of SALMON and its implementation can be found in \cite{SALMON}. The atomic structure calculated by QE was used to determine the ground state in the GS mode of SALMON by solving the Kohn-Sham equation:
	\begin{equation}
		\begin{split}
			\varepsilon_{b,{\bf k}}u^G_{b,{\bf k}}({\bf r})=\Bigg[\frac{1}{2m}{\left(-i\hbar\nabla+\hbar{\bf k}\right)}^{2}\\
			-e\varphi^G({\bf r})+\hat{v}_{{\rm NL}}^{{\bf k}}+{v}^G_{{\rm xc}}({\bf r})\Bigg]u^G_{b,{\bf k}}({\bf r}),
		\end{split}.
		\label{1}
	\end{equation}
	Here, the scalar potential $\varphi({\bf r},t)$ includes the Hartree potential from the electrons and the local part of the ionic pseudopotentials, and $\hat{v}_{{\rm NL}}^{{\bf k}}\equiv e^{-i{\bf k}\cdot{\bf r}}\hat{v}_{{\rm NL}}e^{i{\bf k}\cdot{\bf r}}$. Here, $\hat{v}_{{\rm NL}}$ and ${v}_{{\rm xc}}({\bf r},t)$ are the nonlocal part of the ionic pseudopotentials \cite{TM1993} and the exchange-correlation potential, respectively. The parameters of SALMON are listed in Table \ref{t2}. Although sophisticated exchange-correlation potentials, such as meta-GGA \cite{mBJ,TBmBJ} provide better band gap estimates for silicon, their computational cost is nearly twice that of LDA potentials. Therefore, we used LDA for ${v}_{{\rm xc}}({\bf r},t)$ in this study \cite{PZ1981}.

	The electron dynamics was calculated using the time-dependent Kohn-Sham equation.
	\begin{equation}
		\begin{split}i\hbar\frac{\partial}{\partial t}u_{b,{\bf k}}({\bf r},t)=\Big[\frac{1}{2m}{\left(-i\hbar\nabla+\hbar{\bf k}+\frac{e}{c}{\bf A}(t)\right)}^{2}\\
			-e\varphi({\bf r},t)+\hat{v}_{{\rm NL}}^{{{\bf k}+\frac{e}{\hbar c}{\bf A}(t)}}+{v}_{{\rm xc}}({\bf r},t)\Big]u_{b,{\bf k}}({\bf r},t),
		\end{split}
		\label{2}
	\end{equation}
	where ${\bf A}(t)$ is the applied vector potential.
	The laser field is described by
	\begin{equation}
		{\bf A}(t)=
		\begin{cases}
			{\bf e}_z A_0 \cos^4\left(\frac{t}{T_p}\pi \right)\cos(\omega_0 t) & 0<t<T_p \\
			0& T_p<t< T_e, \label{eq:field}
		\end{cases}
	\end{equation} 
	where $A_0$ and $\omega_0$ denote the amplitude of the vector potential at the peak and the laser frequency, respectively.
	$\hbar\omega_0$ was set to 1.2~eV.
	The pulse length $T_p$ was set to 40fs and the computation was terminated at $T_e=50$~fs.

	\begin{table}
		\caption{List of system and parameters for the Quantum ESPRESSO. "n" represents the number of defects in the cell.}
		\label{t1}
		\begin{tabular}{ccccc}
			\hline 
			\multicolumn{1}{c}{System} & cell & \multicolumn{1}{c}{N$_k$} & \multicolumn{1}{c}{cutoff energy (Ry)} & \\
			\hline
			Si$_{512}$ nO$_i$ &$4\times 4\times 4$ & $2^3$ &80\\ 
			Si$_{512-n}$nV$_{Si}$  &$4\times 4\times 4$ & $2^3$& 120\\ 
			\hline
		\end{tabular}
	\end{table}
	
	\begin{table}
		\caption{List of parameters for SALMON.}
		\label{t2}
		\begin{tabular}{cccc}
			\hline 
			\multicolumn{1}{c}{System}  & \multicolumn{1}{c}{N$_k$} & \multicolumn{1}{c}{N$_r$} &  Cell size (\AA)  \\
			\hline
			Si$_{512}$ nO$_i$ & $\Gamma$ & 88$^3$ &  a=b=c=21.72 \\ 
			Si$_{512-n}$nV$_{Si}$  & $\Gamma$ & 98$^3$ & a=b=c=21.72 \\ 
			c-Si& $16^3$ & 22$^3$ &  a=b=c=5.43 \\
			a-Si (512 atom)& $\Gamma$ & 80$^3$ & a=b=c=21.97 \\
			\hline
		\end{tabular}
	\end{table}

	\section{Results \label{sec:results}}
	\subsection{Interstitial Oxygen}
	
\begin{figure}
		\includegraphics[width=0.5\textwidth]{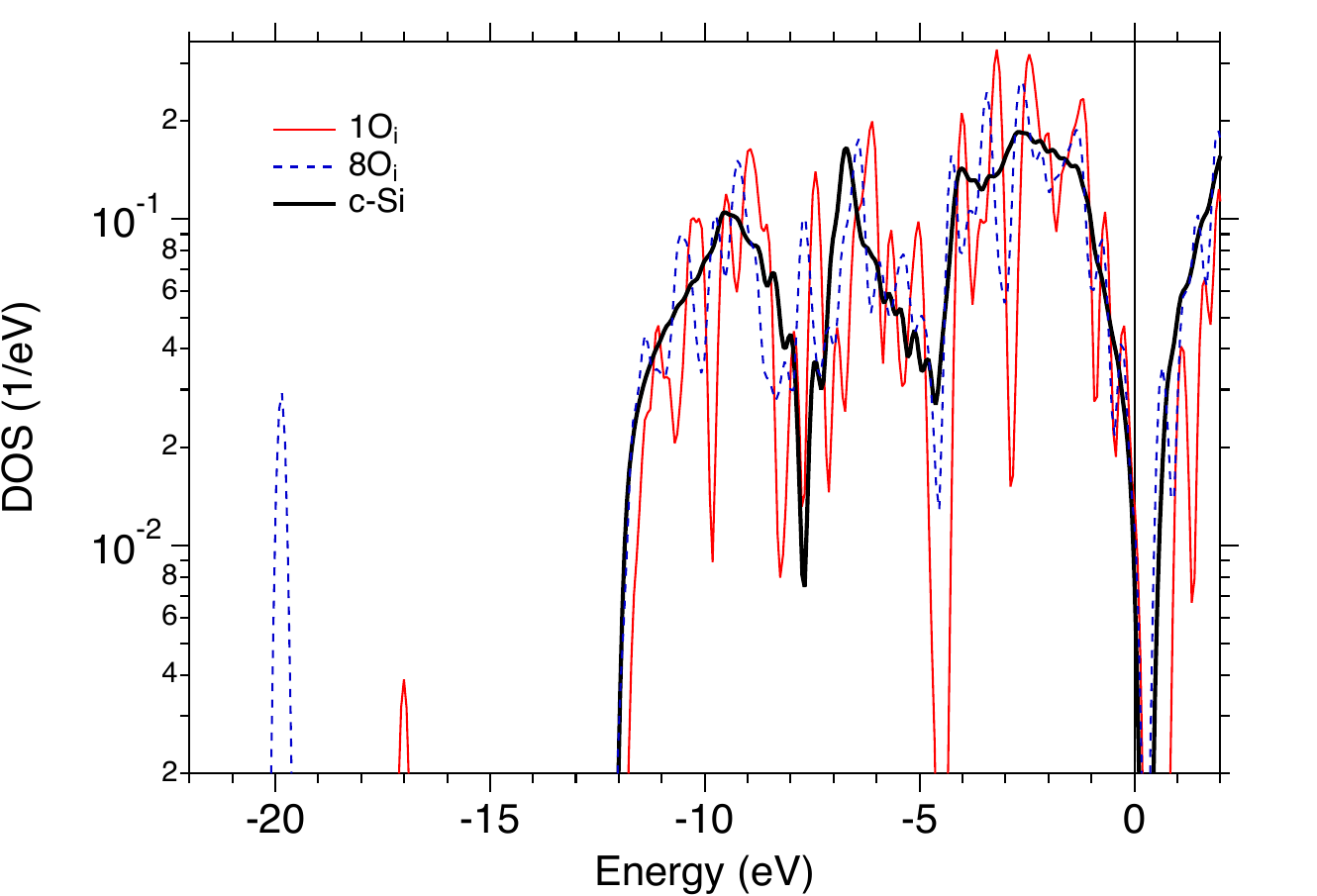}
		\caption{Normalized density of states of 1O$_i$ (red solid line), 8O$_i$ (blue dashed line), and c-Si (black thick line). }
		\label{DOS_Oi}
	\end{figure}

	Figure~\ref{DOS_Oi} compares the normalized density of states (DOS) for 1O$_{i}$ and 8O$_{i}$ with that of crystalline silicon (c-Si). 
	The zero-energy level was set at the top of the occupied energy state. Despite the overall similarity of the DOS structures , a peak associated with oxygen occurs at approximately $-20$~eV. 
	Oxygen atoms attract electrons from silicon, creating deeply bound states around the defects, which do not significantly contribute to the laser excitation.
	
\begin{figure}
		\includegraphics[width=0.5\textwidth]{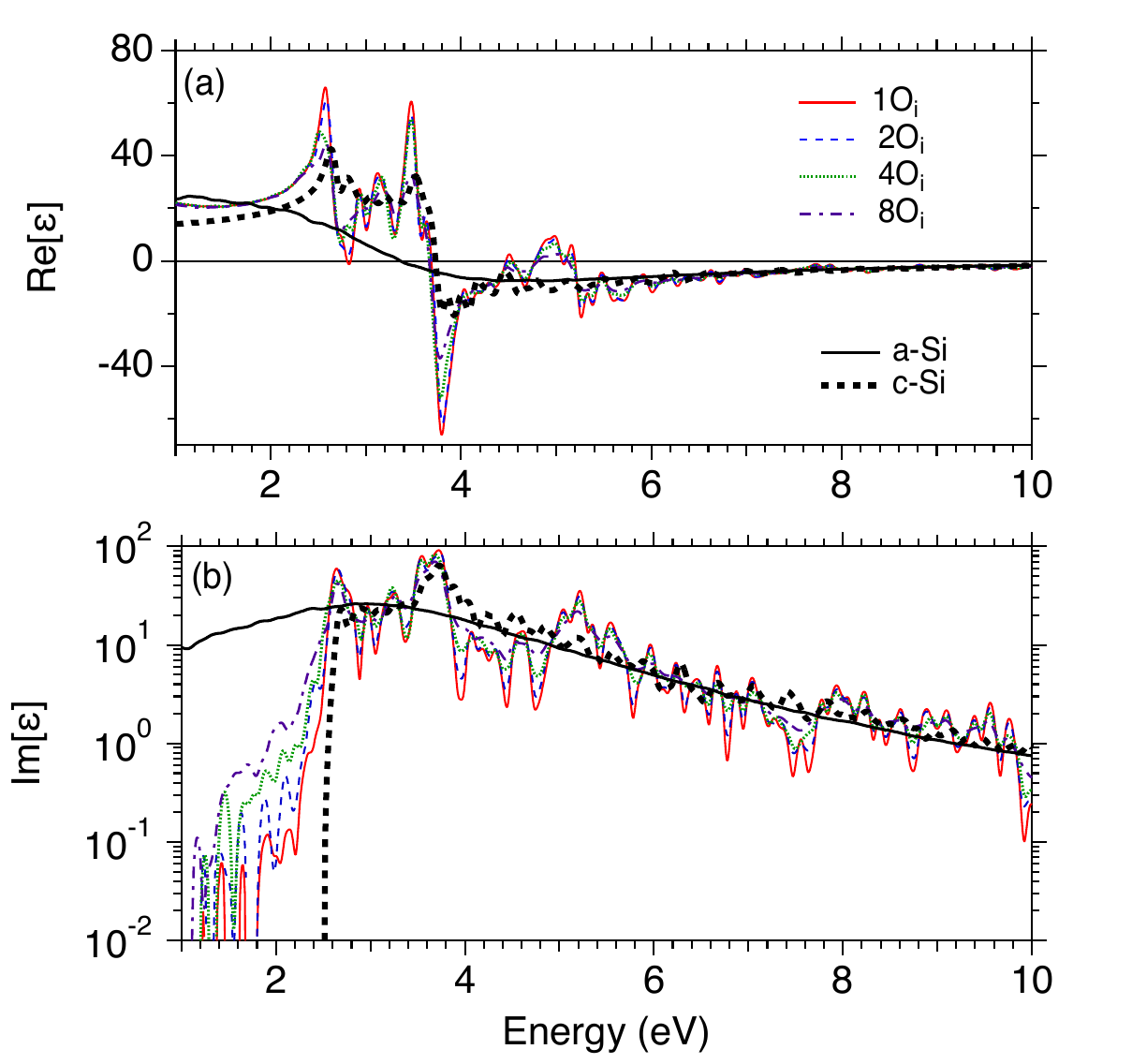}
		\caption{Dielectric function corresponding to various interstitial oxygen densities. c-Si and a-Si are shown as references for the extreme cases. }
		\label{LR_Si_Oi}
	\end{figure}

	Figure~\ref{LR_Si_Oi} shows the dielectric function $\varepsilon(\omega)$. The calculated optical bandgap ($E_o$) is approximately 2.5 eV, which is smaller than the experimental value (3.1 eV) owing to the use of the LDA potential. Variations in $\varepsilon(\omega)$ below $E_o$ are significantly dependent on the oxygen density. While the real part of $\varepsilon(\omega)$ (Fig.\ref{LR_Si_Oi}(a)) exhibits minimal dependence on defect density, several peaks in the imaginary part (Fig.\ref{LR_Si_Oi}(b)) indicate the presence of color center states below $E_o$. Below $E_o$, absorption increases with the oxygen density, approaching the spectrum of amorphous Si (a-Si). This trend suggests a blue shift in the optical bandgap due to the distortion of the crystal structure around the defects, which disrupts the symmetry of the system.
	
\begin{figure}
		\includegraphics[width=0.5\textwidth]{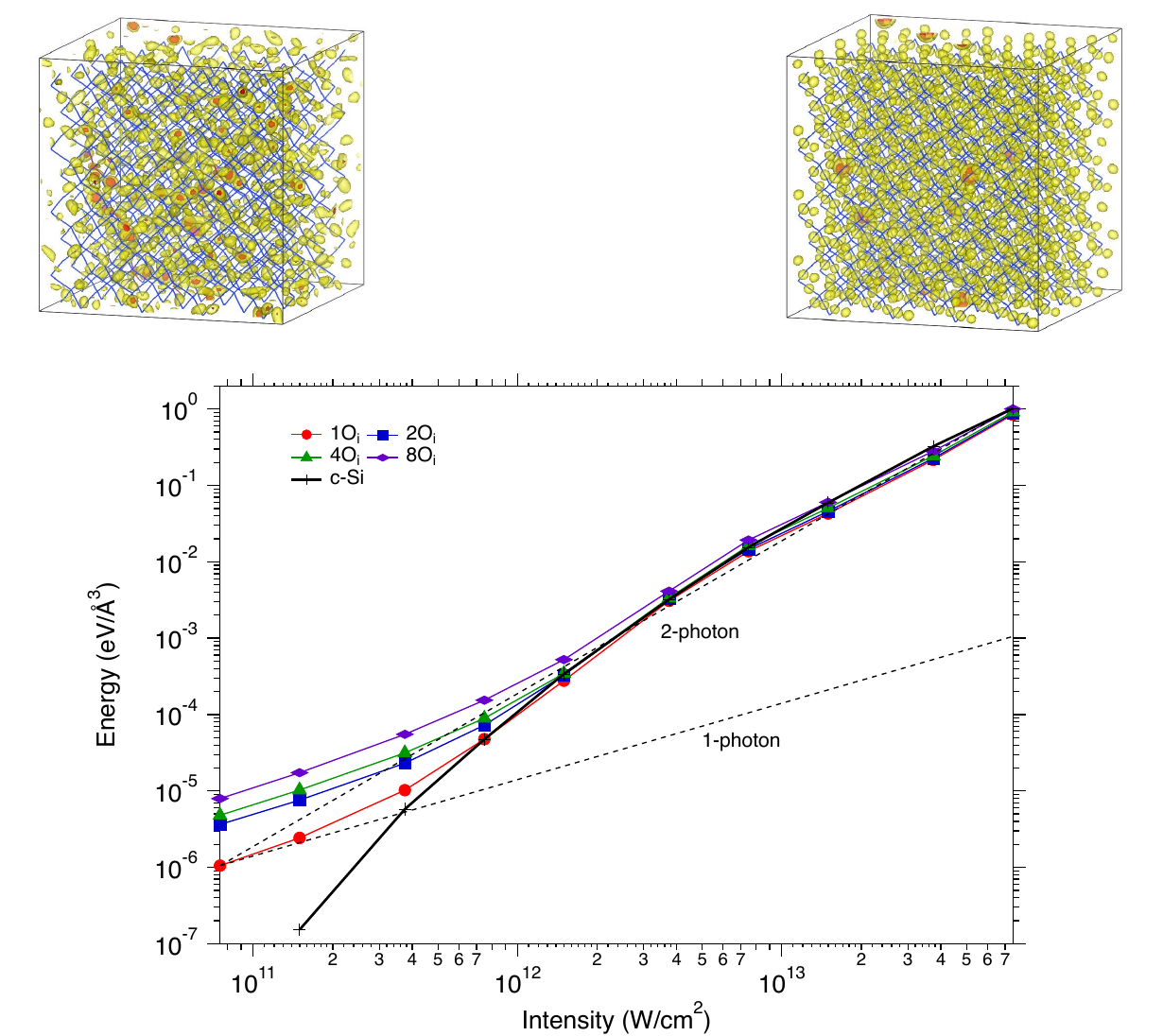}
		\caption{Absorbed energy density as a function of the peak laser intensity. 
			The laser intensity is scaled with respect to the incident intensity using $\rm{Re}[\varepsilon(\omega=1.2eV)]=20$ in 
			Fig.~\ref{LR_Si_Oi}.
			(Insets) Change in the electron density ($\Delta \rho$) from the ground state. The red and yellow isosurfaces indicate $\Delta \rho=-6.8\times 10^{-4}$~\AA$^{-3}$ and $\Delta \rho=-3.4\times 10^{-4}$~\AA$^{-3}$, respectively.}
		\label{In_Dep_Si_Oi}
	\end{figure}

	Figure \ref{In_Dep_Si_Oi} illustrates the dependence of the absorbed energy density on the laser intensity at various oxygen densities. The black solid line with diamonds represents the case of crystalline silicon (c-Si). 
	The pulse length $T_p$ was set to 40 fs, and the computations were terminated at $T_e=50$~fs. The laser intensity was scaled to the intensity ($I$) of the incident pulse by using the following relationship:
	\begin{equation}
		I = \left( \frac{1+n_{Si}}{2} \right)^2 I_m.
	\end{equation}
	Here, $I_m$ is the laser intensity experienced by the electrons and $n_{Si}=\sqrt{20}$ is the refractive index calculated from the real part of $\varepsilon$($\omega=1.2~\rm{eV}$), 
	which remains almost unaffected by defects (Fig. ~\ref{LR_Si_Oi}(a)). The dashed lines indicate the dependence of the one- and two-photon processes. At low intensities, 
	the one-photon process dominated in the presence of  O$_i$. In contrast, c-Si exhibits a three-photon process at approximately $1 \times 10^{12}$~W/cm$^2$. 
	Above $1 \times 10^{13}$~W/cm$^2$, all cases exhibit similar excitation energies. The electron temperature dependence was examined up to 1000~K; however, no significant temperature dependence was observed.

	To gain detailed insight into the excitation process, contour plots of the electron density for 8O$_{i}$ are shown in the insets. The change in electron density ($\Delta \rho$) indicates electron excitation. The red and yellow isosurfaces, which were drawn using VESTA\cite{VESTA2011}, represent $\Delta \rho=-6.8\times 10^{-4}$\AA$^{-3}$ and $\Delta \rho=-3.4\times 10^{-4}$\AA$^{-3}$, respectively. .
	Non-uniform excitation occurs at the lowest intensity (left inset in Fig.\ref{In_Dep_Si_Oi}).  This observation, along with the relatively continuous absorption shown in Fig.\ref{LR_Si_Oi}(b), 
	suggests that the breaking of crystal symmetry due to defects plays a significant role. The excitation becomes more uniform (right inset of Fig. ~\ref{In_Dep_Si_Oi}) as the laser intensity increases. 
	This transition from nonuniform to uniform excitation indicates a shift from a one-photon process associated with defects to a nonlinear process in bulk Si. 
	The Keldysh parameter reached unity at a laser intensity of $7\times10^{14}$~W/cm$^2$, which is beyond the range of our calculations. 
	Therefore, the transition from local to uniform excitation can be attributed to higher-order photoabsorption processes, rather than tunneling.
	
	\subsection{Silicon vacancy}
	
\begin{figure}
		\includegraphics[width=0.5\textwidth]{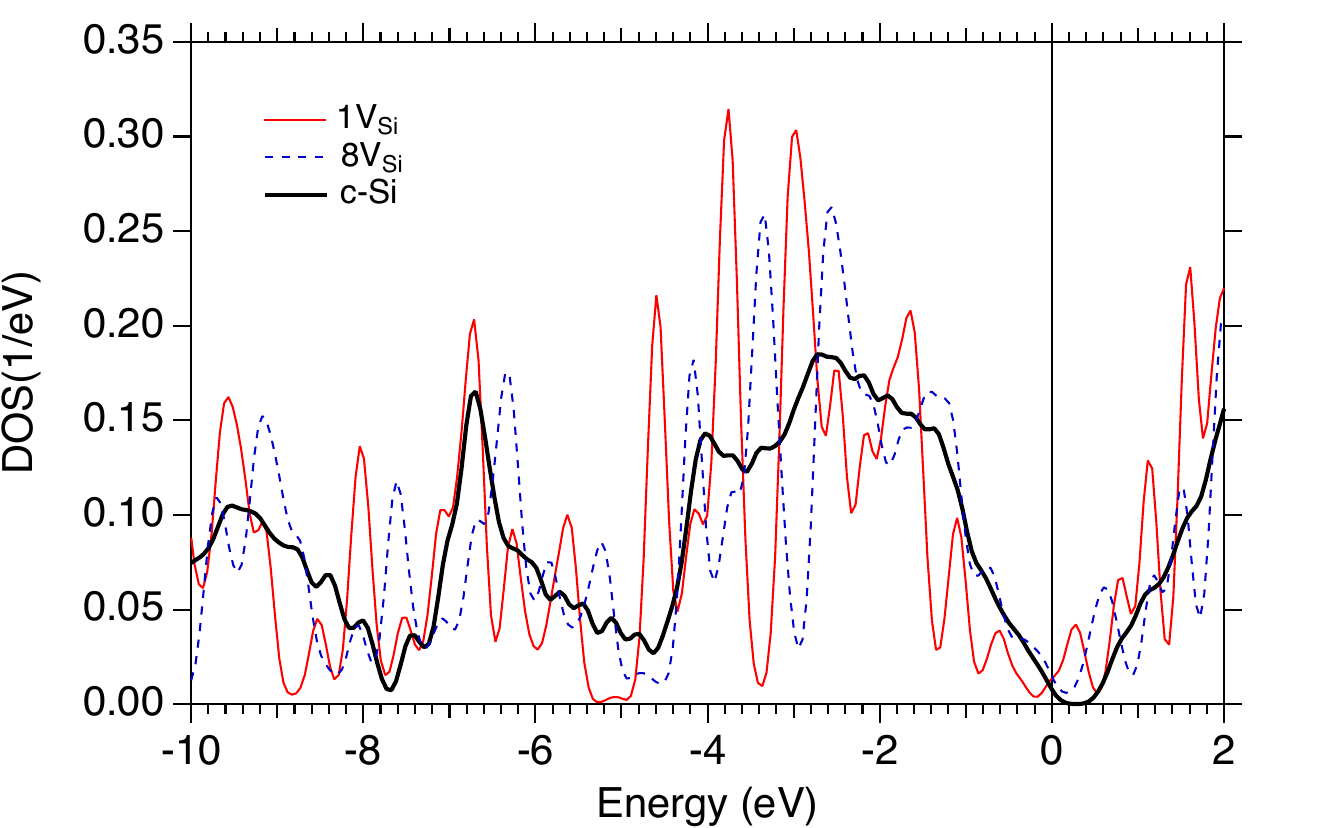}
		\caption{Density of states with silicon vacancies. 0~eV is defined by the highest occupied state.}
		\label{DOS_Vsi}
	\end{figure}

	V$_{Si}$ is a common defect. 
	Although they may exist before laser irradiation they could also be created by the high temperature of silicon after irradiation.
	Figure~\ref{DOS_Vsi} shows the normalized DOS for 1, 8 V$_{Si}$ and c-Si.
	The energy is defined as the energy of the highest occupied energy state.
	Some peaks are observed around 0~eV , which do not exist in O$_i$.
	These peaks imply intense color-centered absorption.
	
\begin{figure}
		\includegraphics[width=0.5\textwidth]{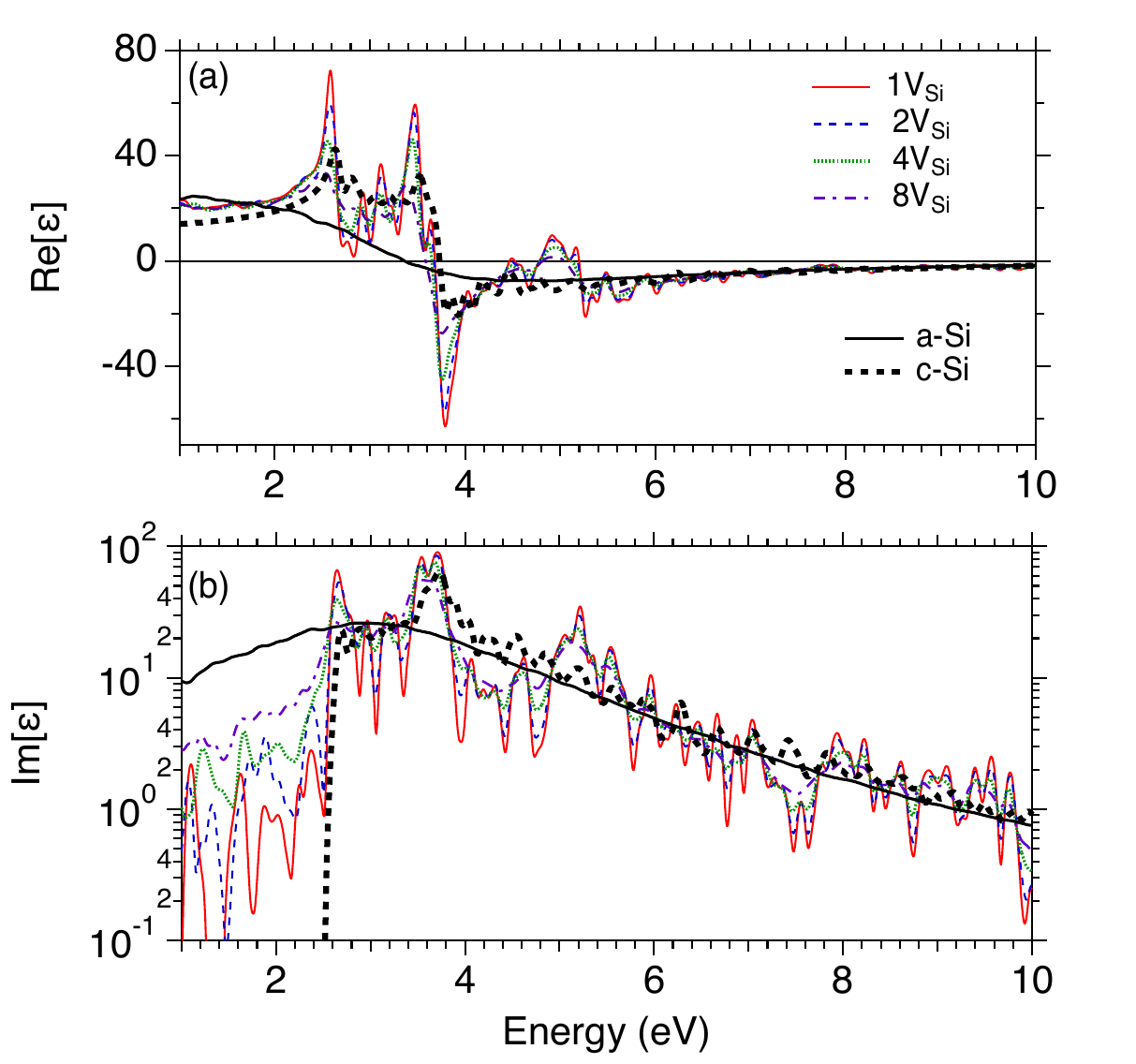}
		\caption{Dielectric functions with silicon vacancies.}
		\label{LR_Vsi}
	\end{figure}

	Figure~\ref{LR_Vsi} illustrates the dielectric function of Si with respect to  V$_{Si}$.
	Although the real part of $\varepsilon$ does not significantly change, 
	the change in Im$[\varepsilon]$ owing to the color centers is more intense than that caused by the O$_{i}$.
	V$_{Si}$ creates dangling bonds and distorts the crystal structure via Jahn–Teller distortion \cite{Sugino92,Iwata2008}. 
	Therefore, as the density of V$_{Si}$ increases, the optical gap exhibits a larger blue shift and intense absorption by the color centers.
	The blue shift is also larger than that due to O$_i$.
	
\begin{figure}
		\includegraphics[width=0.5\textwidth]{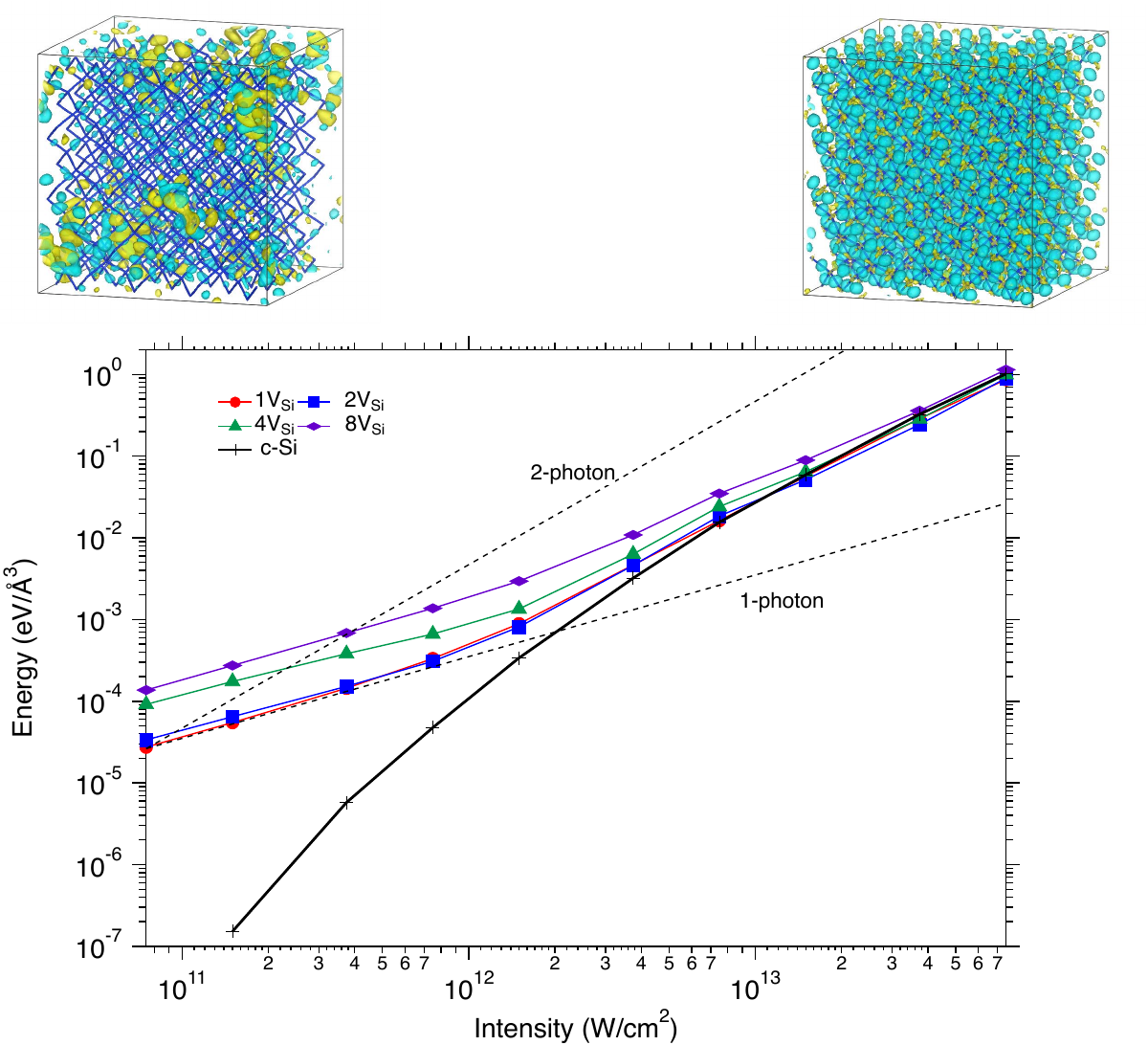}
		\caption{Intensity dependence of the absorbed energy density with V$_{Si}$. (Insets) Change of the electron density from the ground state. The cyan and yellow isosurfaces indicate $\Delta \rho=-0.002$~\AA$^{-3}$and $\Delta \rho=0.002$~\AA$^{-3}$, respectively.}
		\label{In_Dep_Si_V}
	\end{figure}

	Figure \ref{In_Dep_Si_V} shows the dependence of the electron excitation on the laser intensity and V$_{Si}$ density. The one-photon process dominates in all cases below an intensity of $1\times10^{13}$ W/cm$^2$. The energy density does not correlate with the defect density in this regime. Because Im$[\varepsilon]$ in Fig.\ref{LR_Vsi}(b) indicates nearly the same one-photon absorption intensity at 1.2 eV, 1V$_{Si}$ and 2V$_{Si}$ have similar absorbed energies. This result highlights the importance of defect states in laser excitation.

	The insets illustrate $\Delta \rho$ for 8V$_{Si}$ at $7.5\times10^{11}$ W/cm$^2$ (left inset of Fig.\ref{In_Dep_Si_V}, with surfaces of $\pm 0.002$~\AA$^{-3}$) and $7.5\times10^{13}$~W/cm$^2$ (right inset of Fig.\ref{In_Dep_Si_V}, with surfaces of $\pm 0.047$~\AA$^{-3}$). The yellow and cyan surfaces represent positive and negative $\Delta \rho$, respectively. At low intensities, excitation occurs in relatively localized areas compared to the O$_i$ case, suggesting strong absorption by the color centers. At higher intensities, the excitation becomes more uniform, similar to the O$_i$ case, and the excitation energies approach those of c-Si. This indicates that color centers and defect states significantly affect excitation at lower intensities, whereas bulk silicon properties dominate at higher intensities. The linear response calculations (Figs. \ref{LR_Si_Oi} and \ref{LR_Vsi}) show that Im[$\varepsilon$] remains unchanged above $E_o$, which is consistent with the behavior in the high-intensity regime (Figs. \ref{In_Dep_Si_Oi} and \ref{In_Dep_Si_V}). Multiphoton absorption above the optical bandgap is unaffected by defects.
	
	\begin{figure}
		\includegraphics[width=0.5\textwidth]{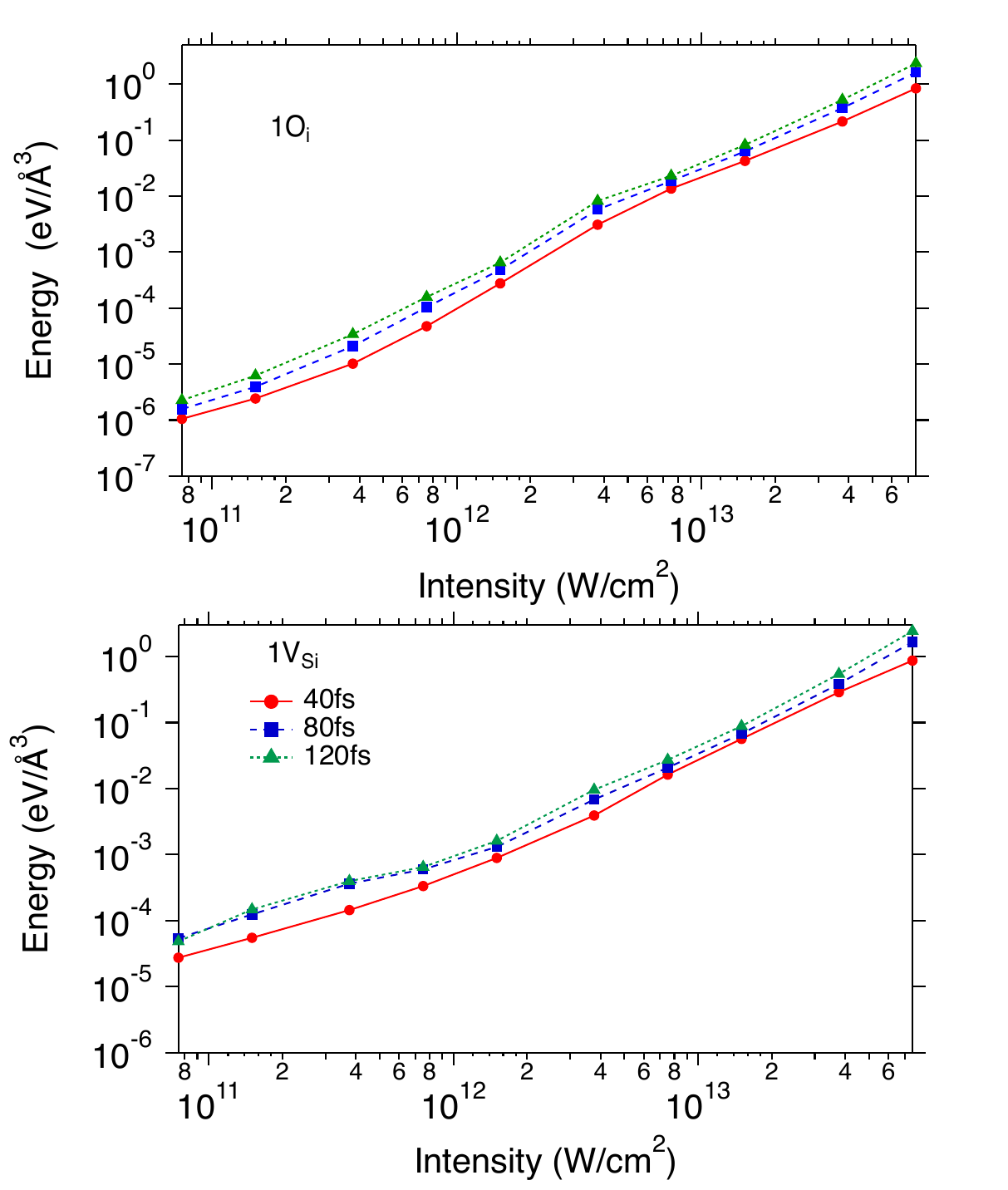}
		\caption{Pulse duration dependence of 1O$_i$ (upper panel) and 1V$_{Si}$ (bottom panel). }
		\label{PD_Vsi}
	\end{figure}
	
	\begin{figure}
		\includegraphics[width=0.5\textwidth]{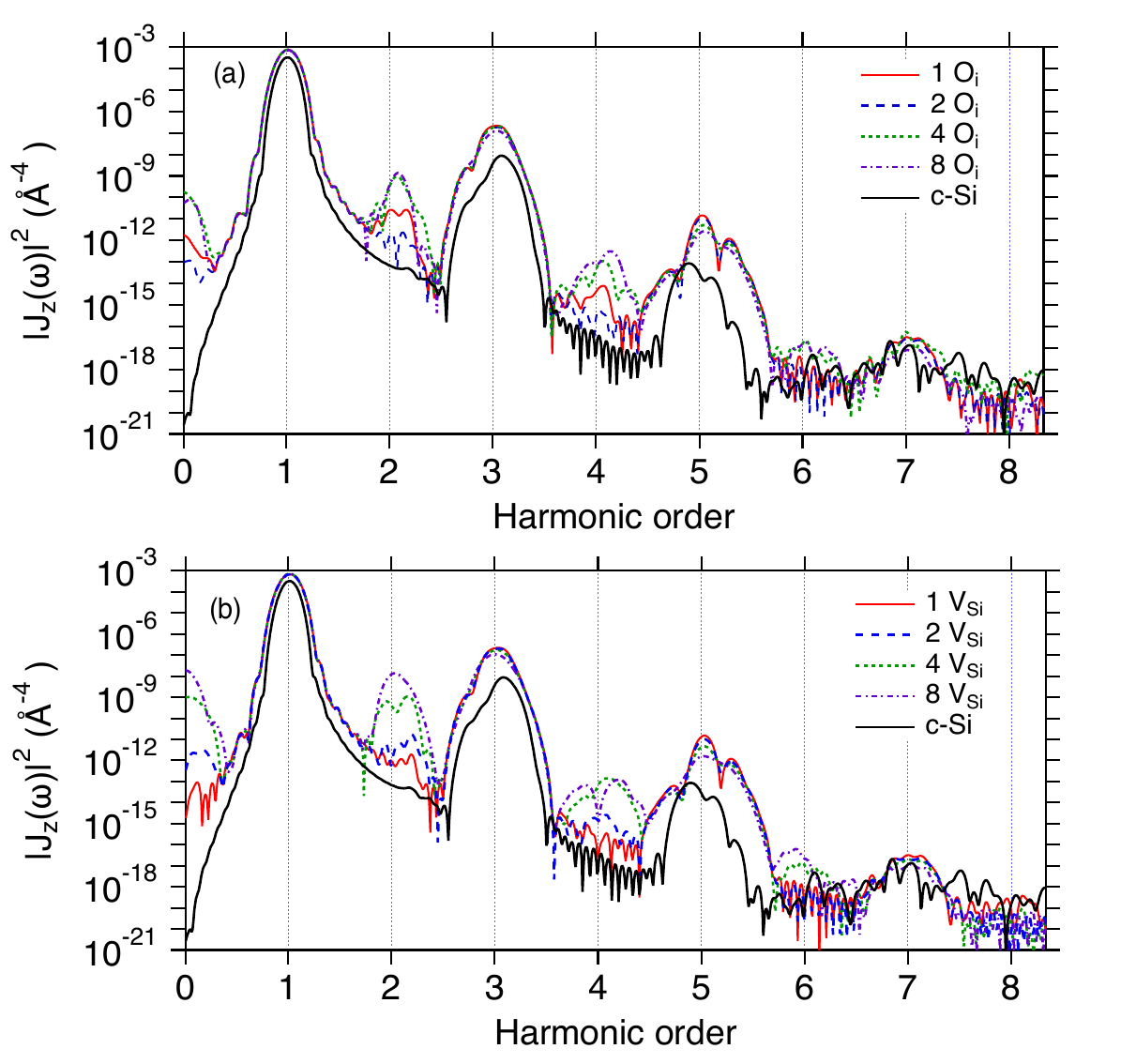}
		\caption{High-harmonic generation with (a) interstitial oxygens and (b) silicon vacancies.The incident laser intensity is assumed to be $7.5\times 10^{11}$~W/cm$^2$.}
		\label{hhg}
	\end{figure}
	\subsection{Defect type dependence}
	
	In this subsection, we explore the dependence of optical properties on defect types in Si, focusing on how color centers and their associated defects influence the photoabsorption efficiency and nonlinear optical responses. 
	
	Defect types, particularly color centers, are critical in determining the optical characteristics of materials. Although these color centers can significantly enhance photoabsorption efficiency, their effectiveness is inherently constrained by their density. When the excitation at a color center reaches saturation, a phenomenon known as saturable absorption occurs, which limits the overall efficiency of the material.
	
	Figure~\ref{PD_Vsi} illustrates the pulse-duration dependence the interstitial oxygen (O$_i$) ( upper panel) and silicon vacancies (V$_{Si}$) ( lower panel). When photoabsorption is unsaturated, the absorbed energy typically increases linearly with pulse duration. However, for O$_i$, this relationship is not strictly linear, although the dependence on the laser intensity remains similar. This suggests a weak saturation of photoabsorption in O$_i$ and indicates that while saturation effects are present, they do not dominate the behavior.
	In contrast, the pulse duration dependence of V$_{Si}$ exhibits characteristics that differ from those of O$_i$. Specifically, the excitation energies of the $T_p=$80~fs and 120~fs pulses remains nearly identical at approximately an intensity of $1\times10^{12}$~W/cm$^2$. This behavior indicates saturable absorption in silicon vacancies at this intensity level. The differences arise from the underlying absorption mechanisms: while interstitial oxygen (1O$_i$) involves relatively weak one-photon absorption necessitating higher-order photon processes, the one-photon absorption process is dominant in silicon vacancies (1V$_{Si}$), leading to a pronounced saturable absorption effect in 1V$_{Si}$.
	
	Next, we discuss how defects influence the nonlinear optical response, particularly in high harmonic generation (HHG). HHG reflects the symmetry of the system. In c-Si, which possesses inversion symmetry, only odd-order harmonics are allowed. However, when defects break this symmetry, even-order harmonics and differential frequency generation (DFG) are possible.
	Figure~\ref{hhg} presents the HHG spectrum under a laser intensity of $7.5\times 10^{10}$~W/cm$^2$ as a function of harmonic order. The vertical axis represents the intensity of the Fourier-transformed electronic current. For both defect types, the intensities of the even-order harmonics and DFG increase as a function of the defect density. The Si vacancies shown in Fig. ~\ref{hhg} (b) demonstrate a significant increase in DFG and second harmonic generation compared to interstitial oxygens, as shown in Fig.~\ref{hhg} (a).
    This result imply that distortion of the crystal structure by silicon vacancy is larger than that by the interstitial oxygen.
	It is noteworthy that the 1O$_i$ defect configuration exhibited more intense even-harmonic generation than the 2O$_i$ configuration did. This result underscores the importance of defect positioning in determining how defects disrupt the symmetry of Si crystals. 
	
	\section{Summary \label{sec:summary}}
	In this study, we used TDDFT to investigate the laser excitation process of silicon with defect states. Interstitial oxygen atoms and silicon vacancies were considered examples of point defects. Our findings revealed significant enhancement in photoabsorption owing to the defect states at low intensity region. However, at higher intensities, the excitation energy was nearly identical to that of defect-free Si. Thus, at higher intensities, the excitation from the valence band to the conduction band of bulk Si via higher-order photon processes dominated, rendering the influence of the defects negligible. Additionally, we observed that the enhancement effect of the defect states was constrained by the pulse length because saturable absorption occurred with relatively weak and long laser pulses.
	Furthermore, we suggest that distortion of the symmetry by defects induces the even-order harmonic-generation and differential frequency generation.
	
	\section{Acknowledgment}
	
	\noindent 

 The authors thank Prof.~Kenichi Ishikawa (The University of Tokyo), and Prof.~Kazuhiro Yabana (University of Tsukuba) for fruitful discussions.
	Funding: This study was supported by MEXT Quantum Leap Flagship Program (MEXT Q-LEAP)
	under Grant No. JPMXS0118067246, and JSPS KAKENHI Grant Number 24K01224. 
	Numerical calculations are performed using the computer facilities of the Fugaku through the HPCI System
	Research Project (Project ID: hp230273).
  This work is partially supported by World-leading Innovative Graduate Study Program Co-designing Future Society (WINGS CFS).
	\bibliography{Ref_Defect}
	
\end{document}